\documentclass[11pt, a4paper]{article}
\pdfoutput=1

\usepackage[height=22cm,width=16cm, centering]{geometry}
\usepackage{setspace}

\usepackage[utf8]{inputenc}

\usepackage{amsmath}
\usepackage{amssymb}
\usepackage{graphicx}
\usepackage{subcaption}
\usepackage{color}
\usepackage[sort&compress, numbers, merge]{natbib}

\definecolor{purple}{rgb}{0.5 ,0, 0.7}
\definecolor{bluegreen}{rgb}{0, 0.45, 0.35}
\definecolor{sakura}{rgb}{1 ,0.52, 0.74}
\definecolor{wakakusa}{rgb}{0.45 ,0.74, 0}
\definecolor{brown}{rgb}{0.48 ,0.23, 0}
\definecolor{skyblue}{rgb}{0.21 ,0.7, 1.}
\definecolor{purplegray}{rgb}{0.35,0.35,0.73}
\usepackage{hyperref}
\hypersetup{colorlinks=true, linkcolor=bluegreen, citecolor=purple, urlcolor=blue}

\def\Mpl{M_{\rm P}}

\allowdisplaybreaks

\begin{document}

\begin{titlepage}
\begin{center}
\leavevmode \\
\vspace{ 0cm}

\hfill KEK-TH-2046\\

\noindent
\vskip 2 cm
 {\Large  Semianalytic Calculation of Gravitational Wave Spectrum \\ \vskip 1mm  Nonlinearly Induced from Primordial Curvature Perturbations}

\vglue .6in

{\Large  
 Kazunori Kohri$^{1,2,3}$ and Takahiro Terada$^1$}

\vskip 1. cm

{\textit {\scriptsize
$^1$Theory Center, IPNS, KEK, 1-1 Oho, Tsukuba, Ibaraki 305-0801, Japan\\
$^2$The Graduate University for Advanced Studies (SOKENDAI), 1-1 Oho, Tsukuba, Ibaraki 305-0801, Japan\\
$^3$Rudolf Peierls Centre for Theoretical Physics, The University of Oxford, 1 Keble Road, Oxford OX1 3NP, United Kingdom
}}

\end{center}

\vglue 0.7in

\begin{abstract}
Whether or not the primordial gravitational wave (GW) produced during inflation is sufficiently strong to be  observable, GWs are necessarily produced from the primordial curvature perturbations in the second order of perturbation.  The induced GWs can be enhanced by curvature perturbations enhanced at small scales or by the presence of matter-dominated stages of the cosmological history. 
We analytically calculate the integral in the expression of the power spectrum of the induced GWs which is a universal part independent of the spectrum of the primordial curvature perturbations. This makes the subsequent numerical integrals significantly easy.  In simple cases, we derive fully analytic formulas for the induced GW spectrum.
\end{abstract}
\end{titlepage}

%%%%%%%%%%%%%%%%%%%%%%%%%%%%%%%%%%%%%%%%%
\section{Introduction}

% Gravitational wave
Gravitational wave (GW) astronomy began after the detection of GWs by the LIGO and Virgo collaborations~\cite{Abbott:2016blz, Abbott:2016nmj, Abbott:2017vtc, Abbott:2017oio, Abbott:2017gyy},
  and more signals or constraints are awaited for current and future precise observations. 
It is now important to study what we can learn about the early Universe or new physics beyond the Standard Model using GWs as probes.
Currently, there is only an upper bound on the strength of the primordial GW~\cite{Grishchuk:1974ny, Starobinsky:1979ty} in terms of the tensor-to-scalar ratio, $r < 0.07$ (95\% confidence level; Planck, BICEP2/Keck Array combined)~\cite{Array:2015xqh} at the pivot scale $k=0.05\, \text{Mpc}^{-1}$, from the cosmic microwave background (CMB) observations. 

% Second order effect
Whether or not the primordial GW is observable, there exits an independent generation mechanism for GWs.\footnote{
Apart from the induced second-order GWs, which are the topic of this paper, there are other mechanisms of GW production in the early Universe, including those associated with preheating~\cite{Khlebnikov:1997di, Easther:2006gt, Easther:2006vd, Dufaux:2007pt}, phase transitions~\cite{Witten:1984rs, Hogan:1986qda, Jinno:2016vai}, and topological defects such as cosmic strings~\cite{Vachaspati:1984gt, Damour:2004kw}.
In particular, it should be noted that the GWs are also produced from the primordial curvature perturbations in the subhorizon when shocks are formed~\cite{Pen:2015qta}. It was reported in Ref.~\cite{Pen:2015qta} that the resultant GW power spectrum is similar to that of the  induced GWs we are considering, but the frequency is lowered by $\mathcal{P}_\zeta^{1/2}$, which would affect the constraints on PBH scenarios which aim to explain the merger rate of binary black holes of around 30 solar masses.  For light PBH scenarios, GWs emitted by Hawking radiation are relevant~\cite{BisnovatyiKogan:2004bk, Anantua:2008am, Dong:2015yjs}.
}  
The GWs are generated from curvature perturbations in the second order of perturbation~\cite{Ananda:2006af} although the tensor and scalar modes are decoupled in the first order of perturbation as is well known.
It is true that the induced second-order GW is suppressed by the square of the curvature perturbations, but it can be sizable and can even become larger than the primordial (first-order) GW if the primordial curvature perturbations are enhanced at small scales compared to the CMB scale~\cite{Ananda:2006af, Bugaev:2009zh,  Alabidi:2012ex} or if the density perturbations grow in a matter-dominated (MD) phase of the Universe~\cite{Mollerach:2003nq, Baumann:2007zm, Assadullahi:2009nf, Alabidi:2013wtp}.
The enhancement of the primordial curvature perturbations at small scales is realized in some models of inflation, and, in particular,  motivates us to explore scenarios to produce primordial black holes (PBHs), which can explain dark matter and/or the binary black hole merger event rate; see, \textit{e.g.}, Refs.~\cite{Kovetz:2017rvv, Kohri:2018qtx, Inomata:2017vxo, Garcia-Bellido:2017aan, Sasaki:2018dmp} and the references therein.  In particular, the induced GWs are used as constraints on the PBH scenarios, and, conversely, the PBH constraints can be recast as the constraints on the induced GWs~\cite{Saito:2008jc, Saito:2009jt, Bugaev:2009zh, Bugaev:2010bb}. Similarly, the induced GWs can put constraints on inflation models which lead to small scale enhancement of perturbations~\cite{Alabidi:2012ex, Alabidi:2013wtp, Orlofsky:2016vbd}. 
Also, MD eras are expected in some scenarios involving heavy particles or oscillating scalar fields like an inflaton.
Entropy production by the decay of such particles/fields at the end of the MD period can dilute unwanted long-lived particles like gravitinos in supersymmetric theories, which otherwise affects the CMB or big-bang nucleosynthesis (BBN).
In this way, studies of induced GWs (as well as the primordial ones) are motivated by cosmology, astrophysics, and particle physics.

% This paper vs literature
To calculate the power spectrum of the GW produced in the radiation-dominated (RD) Universe, we need to do multiple integrals of a highly oscillating function.  Schematically, 
\begin{align*}
\mathcal{P}_h \sim \int \text{d}k \int \text{d}k'  \left(\int \text{d}t f (k, k', t) \right)^2  \mathcal{P}_\zeta (k) \mathcal{P}_\zeta (k'),
\end{align*}
where $\mathcal{P}_h$ ($\mathcal{P}_\zeta$) is the power spectrum of the induced GW (primordial curvature perturbations), $k$ and $k'$ correspond to the momenta of the scalar source modes, $t$ describes the time when the GW is sourced from the scalar modes (using the Green's function method), and $f(k,k',t)$ is some oscillating function.  We will shortly introduce the precise definition.
This can be numerically done, but it is time consuming and sometimes obscures the underlying physics.
An analytic calculation for the time integral was partially done in the pioneering paper~\cite{Ananda:2006af}, and we complete the calculation to obtain a relatively short, useful expression.
An analytic formula was also obtained in Ref.~\cite{Alabidi:2012ex}, but we find it shows the wrong behavior for the contribution from the long-wavelength modes of the density perturbations. 
Once we obtain an analytic formula for the integral within the parentheses above, it is easy to take an oscillation average analytically, so the subsequent (numerical) integration with respect to the wavenumbers $k$ and $k'$ are greatly simplified and the calculation cost is significantly reduced.

In section~\ref{sec:review}, we review the derivation of the scalar-induced GWs basically following the conventions of Ref.~\cite{Inomata:2016rbd}. 
In section~\ref{sec:analytic}, we analytically calculate the time integral, and take its oscillation average.
We consider the cases of both the RD Universe and MD Universe, and we briefly discuss more general cases there and in Appendix~\ref{sec:general}.
For simple examples, we obtain fully analytic formulas for the power spectrum of the induced GWs.
We conclude in section~\ref{sec:conclusion}.  The usage of our formulas is illustrated in Appendix~\ref{sec:comparison}, where they are compared with future observations.
As a byproduct, we derive a new BBN constraint on relativistic degrees of freedom (gravitons, in our case) in Appendix~\ref{sec:BBN}.

%%%%%%%%%%%%%%%%%%%%%%%%%%%%%%%%%%%%%%%%%%%%%%%%%%%
\section{Basic equations}\label{sec:review}

In this section, we review the derivation of the master formula for the second-order GWs.
We basically follow the conventions of Ref.~\cite{Inomata:2016rbd} and extend their results so that we can use them both in a RD era and in a MD era.  See also Refs.~\cite{Ananda:2006af, Baumann:2007zm} for the derivation.

\subsection{Definitions, energy density, and power spectrum}
We choose the longitudinal (conformal Newtonian) gauge, and the metric reads
\begin{align}
\text{d}s^2 = g_{\mu\nu}\text{d}x^\mu \text{d}x^\nu = -a^2(1+2\Phi) \text{d}\eta^2 + a^2 \left( (1-2\Psi)\delta_{ij}+\frac{1}{2}h_{ij} \right) \text{d}x^i \text{d}x^j,
\end{align}
where $\eta$ is the conformal time.
We neglect the vector perturbations, the first-order GWs and the anisotropic stress, and $\Phi = \Psi$ then follows. The effect of the difference $\Phi-\Psi\neq 0$ was studied in Ref.~\cite{Baumann:2007zm}, and it turns out to be small.
With the above normalization, the second-order graviton action is
\begin{align}
S=\frac{\Mpl^2}{32} \int \text{d}\eta \text{d}^3 x a^2  \left( h'_{ij} h'{}_{ij} - h_{ij,k} h_{ij,k} \right),
\end{align}
where $\Mpl=1/\sqrt{8\pi G}=1$ is the reduced Planck mass and primes denote the derivative with respect to the conformal time. The GW energy density $\rho_{\text{GW}}(\eta)=\int \text{d}\ln k \rho_{\text{GW}}(\eta, k)$ can be evaluated in the subhorizon as~\cite{Maggiore:1999vm}
\begin{align}
\rho_{\text{GW}}=\frac{\Mpl^2}{16a^2} \left \langle \overline{h_{ij,k}h_{ij,k}} \right \rangle ,  \label{rho_GW}
\end{align}
 where the overline denotes the oscillation average.  
The Fourier components of the tensor mode are introduced as usual,
\begin{align}
h_{ij} (\eta, {\bf x}) = \int \frac{\text{d}^3 k}{(2 \pi)^{3/2}} \left( e_{ij}^{+} ({\bf k}) h_{{\bf k}}^{+} (\eta) + e_{i j}^{\times} ({\bf k}) h_{{\bf k}}^{\times} (\eta) \right) e^{i {\bf k}\cdot {\bf x}},
\end{align}
where the transverse traceless polarization tensors are defined as $e_{ij}^+ ({\bf k}) = \frac{1}{\sqrt{2}}(e_i ({\bf k}) e_j ({\bf k}) - \bar{e}_i ({\bf k}) \bar{e}_j({\bf k}))$ and $e_{ij}^\times ({\bf k}) = \frac{1}{\sqrt{2}}(e_i ({\bf k}) \bar{e}_j ({\bf k}) + \bar{e}_i ({\bf k}) e_j({\bf k}))$, with $e_i ({\bf k})$ and $\bar{e}_i ({\bf k})$ being normalized vectors orthogonal to each other and to ${\bf k}$.
The dimensionless power spectrum is defined by
\begin{align}
\langle h_{{\bf k}}^{\lambda} (\eta) h_{{\bf k}'}^{\lambda'} (\eta) \rangle =& \delta_{\lambda \lambda'} \delta^3 ({\bf k} + {\bf k}')  \frac{2 \pi^2}{k^3} \mathcal{P}_h (\eta, k),
\end{align}
where $\lambda, \lambda' = + , \times$ represents the polarization index, which we omit in the following.  
We consider parity invariant situations so that both polarizations give the same result.
 The fraction of the GW energy density per logarithmic wavelength is
\begin{align}
\Omega_{\text{GW}}(\eta, k)=\frac{\rho_{\text{GW}}(\eta,k)}{\rho_{\text{tot}}(\eta)}= \frac{1}{24} \left( \frac{k}{a(\eta)H(\eta)} \right)^2 \overline{\mathcal{P}_h(\eta, k)}, \label{Omega_GW}
\end{align}
where we have summed over the two polarization modes.
This $\Omega_{\text{GW}}$ is the observationally relevant quantity, and below we compute the power spectrum $\mathcal{P}_h$.

\subsection{Equations of motion}
The equation of motion for the tensor mode $h_{\bf k}(\eta)$ can be derived straightforwardly from the tensor part of the Einstein equation.  In the second-order equation, squares of first-order quantities also appear.  The first-order perturbations of the energy-momentum tensor can be related to the derivative of the gravitational potential $\Phi$ via the first-order Einstein equation.  Thus, one obtains the tensor equation of motion sourced by the scalar perturbations $\Phi$,
\begin{align}
h''_{\bf k}(\eta) + 2 \mathcal{H} h'_{\bf k}(\eta)+ k^2 h_{\bf k}(\eta) =& 4 S_{\bf k}(\eta),  \label{EOM_h}
\end{align}
where $\mathcal{H}=a H$ is the conformal Hubble parameter, and the source term is given by
\begin{align}
S_{\bf k} =& \int \frac{\text{d}^3 q}{(2 \pi)^{3/2}} e_{ij}({\bf k}) q_i q_j \left( 2\Phi_{\bf q}  \Phi_{{\bf k}-{\bf q}} + \frac{4}{3(1+w)} \left( \mathcal{H}^{-1} \Phi'_{\bf q} + \Phi_{\bf q}\right) \left( \mathcal{H}^{-1} \Phi'_{{\bf k}-{\bf q}} + \Phi_{{\bf k}-{\bf q}} \right)  \right).
\end{align}
We have used $-2 \dot{H} = \rho + P =( 1+w) \rho = 3 (1+w)H^2$, where $w = P/\rho$ is the equation-of-state parameter with $P$ and $\rho$ denoting pressure and energy density.
The Fourier components $\Phi_{\bf k}$ of the gravitational potential are defined similarly to those of the tensor mode (of course without the polarization tensor).
We adopt the Green's function method to solve $h_{\bf k}(\eta)$,
\begin{align}
a(\eta) h_{\bf k}(\eta) = 4 \int^\eta \text{d}\bar{\eta} G_{\bf k}(\eta, \bar{\eta}) a(\bar{\eta}) S_{\bf k}(\bar{\eta}),
\end{align}
where the Green's function $G_{\bf k}(\eta, \bar{\eta})$ is the solution of
\begin{align}
G_{\bf k}''(\eta, \bar{\eta}) +\left( k^2 - \frac{ a''(\eta)}{a(\eta)}\right) G_{\bf k}(\eta, \bar{\eta}) = \delta (\eta - \bar{\eta}). \label{EOM_Green}
\end{align}
The derivatives are with respect to $\eta$.

We need to know the time evolution of the source term $S_{\bf k}(\eta)$.
The gravitational potential obeys the following equation of motion (see \textit{e.g.}~Ref.~\cite{Mukhanov:2005sc}):
\begin{align}
\Phi''_{\bf k} + 3 \mathcal{H} (1 + c_{\text{s}}^2) \Phi'_{\bf k} + (2 \mathcal{H}'+(1+3 c_{\text{s}}^2)\mathcal{H}^2 +c_{\text{s}}^2 k^2) \Phi_{\bf k}  = \frac{a^2}{2} \tau \delta S,  \label{EOM_Phi_complete}
\end{align}
where $c_{\text{s}}^2$ and $\tau$ are defined as $\delta P = c_{\text{s}}^2 \delta \rho + \tau \delta S$, with $S$ being entropy.  In the absence of entropy perturbations and using $c_{\text{s}}^2 = w$, the above equation reduces to
\begin{align}
\Phi''_{\bf k}(\eta) + \frac{6(1+w)}{(1+3w)\eta } \Phi'_{\bf k}(\eta) + w k^2 \Phi_{\bf k}(\eta)=0. \label{EOM_Phi}
\end{align}
In the following, we pull out the primordial value $\phi_{\bf k}$ from the definition of $\Phi_{\bf k} = \Phi(k \eta) \phi_{\bf k}$ so that the transfer function $\Phi (k \eta)$ approaches unity well before the horizon entry.
The primordial value is related to the curvature perturbation as 
\begin{align}
\langle \phi_{\bf k} \phi_{\bf k'} \rangle = \delta ({\bf k}+{\bf k}' ) \frac{2\pi^2}{k^3} \left( \frac{3+3w}{5+3w} \right)^2 \mathcal{P}_\zeta (k),
\end{align}
where $w$ should be evaluated at time well before the horizon entry.
As the above equation implies, we define the ``primordial'' value $\phi_{\bf k}$ as being well before the horizon entry but not too early so that the equation of state of the Universe at the ``primordial time'' is the same as that at the horizon entry. 

One can compute the correlation function $\langle S_{\bf k} (\eta) S_{\bf k'}(\eta') \rangle$ by neglecting the non-Gaussianity of the primordial curvature perturbations.  It involves integration with respect to the wavenumber $\tilde{k}$ corresponding to that of the scalar source $\Phi_{\tilde{\bf k}}$.
It turns out to be useful to introduce the dimensionless variables $u=|{\bf k}-\tilde{\bf k}|/k$ and $v= \tilde{k}/k$.
The details for this calculation can be found in Refs.~\cite{Ananda:2006af, Baumann:2007zm,  Inomata:2017vxo}.
After some algebra, by comparing $\langle S_{\bf k} (\eta) S_{\bf k'}(\eta') \rangle$ with the definition of $\mathcal{P}_h$, one can extract the power spectrum $\mathcal{P}_h$,
\begin{align}
\mathcal{P}_h (\eta, k) =  4  
 \int_0^\infty \text{d}v \int_{\left| 1-v \right |}^{1+v}\text{d} u \left( \frac{4v^2 - (1+v^2-u^2)^2}{4vu} \right)^2 I^2 (v,u,x) \mathcal{P}_\zeta ( k v ) \mathcal{P}_\zeta ( k u ), \label{P_h}
\end{align}
where the dimensionless combination $x\equiv k \eta$ should not be confused with the spatial coordinate.
The function $I(v,u,x)$ is defined as
\begin{align}
I(v,u,x)= \int_0^x \text{d}\bar{x} \frac{a (\bar{\eta})}{a(\eta)} k G_k (\eta, \bar{\eta}) f (v,u,\bar{x}),   \label{I}
\end{align}
and the source information is contained in 
\begin{align}
f (v ,u ,\bar{x}) = & \frac{6(w+1)}{3w+5}\Phi(v\bar{x})\Phi(u\bar{x})+\frac{6(1+3w)(w+1)}{(3w+5)^2} \left( \bar{x}\partial_{\bar{\eta}}\Phi(v\bar{x})\Phi(u\bar{x})+\bar{x}\partial_{\bar{\eta}} \Phi(u\bar{x})\Phi(v\bar{x}) \right) \nonumber \\
& +\frac{3(1+3w)^2(1+w)}{(3w+5)^2}  \bar{x}^2 \partial_{\bar{\eta}}\Phi(v\bar{x})\partial_{\bar{\eta}}\Phi(u\bar{x}),
\end{align}
where $\bar{x} = k \bar{\eta}$, and we have used $\mathcal{H}=aH=2/((1+3w)\eta)$.
Note that the integral defining $\mathcal{P}_h$ includes the square of a single function $I(v,u,x)$.
The integral has been recast in this form by noticing symmetries of the integrand under changes of variables as explained in Ref.~\cite{Inomata:2017vxo}.  This was seemingly unnoticed in the original paper~\cite{Ananda:2006af}, which makes their analytic expression so complicated.
Related to this, both the integrand and the integral region are symmetric under the exchange of $u$ and $v$.
Taking advantage of the above form, we will calculate the function $I(v,u,x)$ analytically in the following section.

%%%%%%%%%%%%%%%%%%%%%%%%%%%%%%%%%%%%%%%%%%%%%%%%%%%
\section{Analytic integration}\label{sec:analytic}

We find it useful to introduce new variables 
$t = u+ v-1$  and 
$s= u-v$ to finally execute the remaining integrals.  
 The Jacobian for this transformation is $1/2$.
Using these variables, the power spectrum is rewritten as
\begin{align}
\mathcal{P}_h (\eta, k) =  2 
 \int_0^\infty \text{d}t \int_{-1}^{1}\text{d} s \left [ \frac{t(2+t)(s^2-1)}{(1-s+t)(1+s+t)} \right ]^2 I^2 (v,u,x) \mathcal{P}_\zeta ( k v ) \mathcal{P}_\zeta ( k u ),  \label{P_h_ts}
\end{align}
where $u=(t+s+1)/2$ and $v=(t-s+1)/2$, and we remind the reader of the definition $x = k \eta$.
In the following, we give expressions of $I$ in terms of $u$ and $v$ as well as $t$ and $s$.
We separately study the cases of a pure RD era and a pure MD era, finally discussing more realistic situations.

%%%%%%%%%%%%%%%%%%%%%%%%%%%%%%
\subsection{Radiation-dominated Universe}\label{ssec:RD}
In the RD Universe, the solution to eq.~\eqref{EOM_Green} for the Green's function of GW is
\begin{align}
k G_{\bf k} (\eta , \bar{\eta}) =& - x \bar{x} (j_0 (x) y_0 (\bar{x}) - y_0 (x) j_0 (\bar{x})) \nonumber \\
=& \sin (x- \bar{x}),
\end{align} 
where $j_0(x)$ ($y_0(x)$) is the spherical Bessel function of the first (second) kind.
The solution to eq.~\eqref{EOM_Phi} for the gravitational potential which approaches $1$ in the past $(x \to 0)$ is
\begin{align}
\Phi(x) = \frac{9}{x^2} \left( \frac{\sin(x/\sqrt{3})}{x/\sqrt{3}} -\cos (x/\sqrt{3}) \right).
\end{align}
The factor $1/\sqrt{3}$ is the sound speed in the RD era.
The gravitational potential decays like $x^{-2}$ at large $x$.

The source function $f$ in the RD era is
\begin{align}
f_{\text{RD}}(v,u,x) =& \frac{12}{u^3 v^3 x^6} \left( 18 u v x^2 \cos \frac{u x}{\sqrt{3}} \cos  \frac{v x}{\sqrt{3}} +(54 - 6 (u^2 +v^2)x^2 + u^2 v^2 x^4 ) \sin \frac{u x}{\sqrt{3}} \sin \frac{v x}{\sqrt{3}}      \right. \nonumber \\
& \qquad \left. + 2\sqrt{3} ux (v^2 x^2 -9) \cos \frac{u x}{\sqrt{3}}  \sin \frac{v x}{\sqrt{3}} + 2\sqrt{3} vx (u^2 x^2 -9) \sin \frac{u x}{\sqrt{3}}  \cos \frac{v x}{\sqrt{3}}  \right) .
\end{align}
This is equal to 4/3 at $x=0$ and decays like $\sim 1/(uvx^2)$ at large $x$.
The factor $a(\bar{\eta})/a(\eta)$ in the definition of $I(v,u,x)$ is equal to $\bar{x}/x$ in the RD era.\footnote{\label{fn:dof}
Precisely speaking, it involves the effective numbers of relativistic degrees of freedom, 
\begin{align}
\frac{a(\bar{\eta})}{a(\eta)} = &  \frac{\bar{\eta}}{\eta} \left( \frac{g_{*}(T(\bar \eta ))}{g_{*}(T(\eta))} \right)^{1/2} \left( \frac{g_{*, s}(T( \eta ))}{g_{*, s}(T( \bar \eta))}  \right)^{2/3}.
\end{align}
Before recombination, both numbers are the same, $g_{*}(T)=g_{*,s}(T)$, and the power is only $1/6$. We neglect these factors for analytic calculations.
} 

Combining this information, we calculate the integral $I(v,u,x)$.
To this end, multiple usages of the trigonometric addition theorem and integration by parts are required~\cite{Ananda:2006af}.
The result is 
\begin{align}
I_{\text{RD}} (v,u,x) =& \frac{3}{4 u^3 v^3 x} \left( -\frac{4}{x^3} \left(  uv (u^2 + v^2 -3) x^3 \sin x - 6 u v x^2 \cos \frac{ux}{\sqrt{3}} \cos\frac{vx}{\sqrt{3}} \right. \right. \nonumber \\
&\left.+ 6\sqrt{3} ux \cos \frac{ux}{\sqrt{3}} \sin \frac{vx}{\sqrt{3}} + 6 \sqrt{3} v x \sin \frac{ux}{\sqrt{3}}  \cos \frac{vx}{\sqrt{3}} -3(6+(u^2+v^2-3)x^2) \sin\frac{ux}{\sqrt{3}}  \sin\frac{vx}{\sqrt{3}}   \right) \nonumber \\
& +(u^2+v^2-3)^2 \left( \sin x \left( \text{Ci}\left( \left( 1-\frac{v-u}{\sqrt{3}} \right)x \right) +\text{Ci}\left( \left( 1+\frac{v-u}{\sqrt{3}} \right)x \right) \right. \right. \nonumber \\
& \left. -\text{Ci}\left( \left| 1-\frac{v+u}{\sqrt{3}} \right|x \right) -\text{Ci}\left( \left( 1+\frac{v+u}{\sqrt{3}} \right)x \right) + \log \left|\frac{3-(u+v)^2}{3-(u-v)^2} \right| \right) \nonumber \\
& +\cos x \left( -\text{Si}\left( \left( 1-\frac{v-u}{\sqrt{3}} \right)x \right) -\text{Si}\left( \left( 1+\frac{v-u}{\sqrt{3}} \right)x \right)  \right. \nonumber \\
& \left. \left. + \text{Si}\left( \left( 1-\frac{v+u}{\sqrt{3}} \right)x \right) +\text{Si}\left( \left( 1+\frac{v+u}{\sqrt{3}} \right)x \right)   \right)  \right), \label{I_RD}
\end{align}
where $\text{Si}$ and $\text{Ci}$ functions are defined as follows: 
\begin{align}
\text{Si}(x) =& \int^x_0 \text{d}\bar{x} \frac{\sin\bar{x}}{\bar{x}}, &
\text{Ci}(x) =& - \int^\infty_x \text{d}\bar{x} \frac{\cos\bar{x}}{\bar{x}}.
\end{align}
We have used the fact that
\begin{align}
\int_0^x \text{d}\bar{x} \frac{\cos A\bar{x} - \cos B \bar{x}}{\bar{x}} = \text{Ci}(Ax)-\log (Ax)-\text{Ci}(Bx) +\log (Bx).
\end{align}
For small $x$, the leading term is independent of $u$ and $v$, $I_{\text{RD}}(v,u,x) \simeq x^2 /2$.

We are interested in the GW spectrum observed today, so let us take the late-time limit $\eta \to \infty$ or $x \gg 1$: 
\begin{align}
I_{\text{RD}}(v,u,x\to \infty) =& \frac{3(u^2+v^2-3)}{4 u^3 v^3 x} \left( \sin x \left( -4uv+(u^2+v^2-3) \log \left| \frac{3-(u+v)^2}{3-(u-v)^2} \right| \right) \right. \nonumber \\
& \left. \qquad \qquad \qquad \qquad - \pi (u^2+v^2-3)\Theta ( v+u-\sqrt{3}) \cos x   \right). \label{I_RD_late-time}
\end{align}
We have used $\lim_{x\to \pm \infty} \text{Si}(x)=\pm \pi/2$ and $\lim_{x\to+\infty}\text{Ci}(x)=0$, and the sign change of the limit of Si is the origin of the Heaviside theta function $\Theta$ in the above expression.
We can see that it redshifts like $x^{-1} \propto a^{-1}$ in this limit.
What we want to know is its oscillation average [see eq.~\eqref{Omega_GW}].
It is
\begin{align}
\overline{I_{\text{RD}}^2(v,u,x\to \infty)} =& \frac{1}{2} \left( \frac{3(u^2+v^2-3)}{4 u^3 v^3 x} \right)^2 \left( \left( -4uv+(u^2+v^2-3) \log \left| \frac{3-(u+v)^2}{3-(u-v)^2} \right| \right)^2  \right. \nonumber \\
&  \left. \qquad \qquad \qquad \qquad \qquad   + \pi^2 (u^2+v^2-3)^2 \Theta ( v+u-\sqrt{3}) \right). \label{I_RD_osc_ave}
\end{align}
In terms of the variables $t=u+v-1$ and $s=u-v$, 
\begin{align}
\overline{I_{\text{RD}}^2(t,s,x\to \infty)} =& \frac{288(-5+s^2+t(2+t))^2}{x^2 (1-s+t)^6 (1+s+t)^6}\left( \frac{\pi^2}{4} (-5+s^2+t(2+t))^2 \Theta (t -(\sqrt{3}-1)) \right. \nonumber \\
& \left.+ \left( -(t-s+1)(t+s+1) + \frac{1}{2} (-5+s^2+t(2+t)) \log \left| \frac{-2+t(2+t)}{3 - s^2} \right| \right)^2 \right). \label{I_RD_osc_ave_ts}
\end{align} 
These formulas are our main results.

Let us discuss some simple examples.
\paragraph{Example 1: Monochromatic source}
Consider the monochromatic curvature perturbations,
\begin{align}
\mathcal{P}_\zeta (k) = A_\zeta \delta (\log k/k_* ),  \label{P_zeta_delta}
\end{align}
where $A_\zeta$ is the overall normalization and $k_*$ is the wavenumber at which the power spectrum has a delta-function peak. 
This may be regarded as a rough approximation of a spectrum with a sharp peak.
For example, $k_*$ should be about $3.5 \times 10^5 \, \text{Mpc}^{-1}$ or $2.7 \times 10^{-14} \, \text{Mpc}^{-1}$ for PBHs produced in a RD era to explain dark matter abundance or the LIGO/Virgo binary black hole merger rate, respectively.

\begin{figure}[tbh]
 \centering
{\includegraphics[width=0.65\columnwidth]{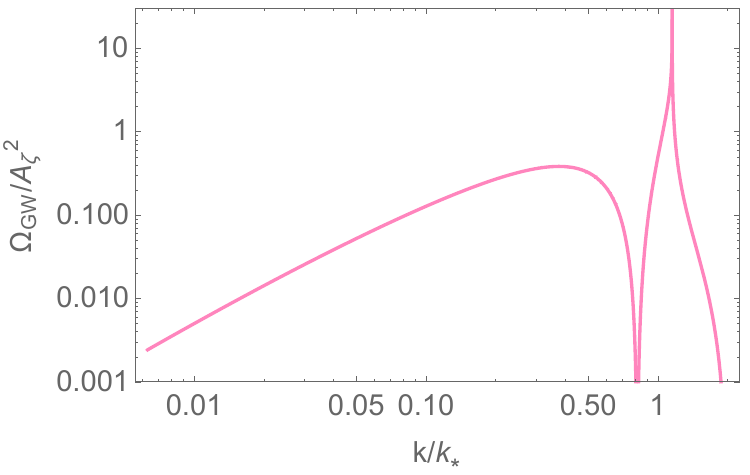}}
  \caption{The energy density fraction $\Omega_{\text{GW}}$ of GWs produced in the RD era, eq.~\eqref{Omega_GW_RD_delta}, from the monochromatic source, eq.~\eqref{P_zeta_delta}.}
 \label{fig:Omega_GW_RD_delta}
  \end{figure}

In this monochromatic case, the GW strength is
\begin{align}
\Omega_{\text{GW}}(\eta, k)=& \frac{3 A_\zeta^2}{64}  \left(\frac{4-\tilde{k}^2}{4} \right)^2  \tilde{k}^2 \left(3 \tilde{k}^2-2\right)^2 \nonumber \\
&  \times  \left( \pi^2 (3 \tilde{k}^2-2)^2 \Theta (2\sqrt{3}-3 \tilde{k}) + \left( 4+(3\tilde{k}^2-2) \log \left| 1- \frac{4}{3 \tilde{k}^2} \right| \right)^2 \right) \Theta (2-\tilde{k}), \label{Omega_GW_RD_delta}
\end{align}
where the dimensionless wavenumber $\tilde{k}\equiv k/k_*$ is introduced for notational simplicity.
The result of Ref.~\cite{Inomata:2016rbd} is reproduced in the small $\tilde{k}$ limit where their approximation is valid.
The logarithmic singularity at $k=(2/\sqrt{3})k_*$  ($u+v = \sqrt{3}$) is due to resonant amplification: the frequency of the source term oscillation is twice that of the gravitational potential $2 \times k_* / \sqrt{3}$. The factor $2$ appears because this is the second-order effect, and $1/\sqrt{3}$ is the ratio between the propagating speeds of GWs and radiation. 
The spectrum vanishes above $k = 2 k_*$ because there are no solutions satisfying the energy and momentum conservation.
Equation.~\eqref{Omega_GW_RD_delta} is shown in Figure~\ref{fig:Omega_GW_RD_delta}.

\paragraph{Example 2: Scale-invariant case}
The scale-invariant power spectrum is
\begin{align}
\mathcal{P}_\zeta (k)= A_\zeta,
\end{align}
where $A_\zeta$ is independent of $k$. We can do a numerical integration to obtain
\begin{align}
\mathcal{P}_h (\eta, k) \simeq & \frac{19.73}{(k\eta)^2} A_\zeta^2, & % 19.733845641738654
\Omega_{\text{GW}}(\eta,k) \simeq & 0.8222  A_\zeta^2,  % 0.8222435684057772 
\label{Omega_GW_RD_SI}
\end{align}
where we have used $\mathcal{H}=\eta^{-1}$ in the RD era.

\paragraph{Example 3: power-law spectrum}
We extend the previous case to a general power-law spectrum,
\begin{align}
\mathcal{P}_\zeta  = A_\zeta \left( \frac{k}{k_*} \right)^{n_{\text{s}}-1},
\end{align}
where $k_*$ is a reference scale and $n_{\text{s}}-1$ controls the spectral tilt.
In this case,
\begin{align}
\mathcal{P}_h (\eta, k) =& \frac{24 Q(n_{\text{s}})}{(k\eta)^2} A_\zeta^2 \left( \frac{k}{k_*} \right)^{2(n_{\text{s}}-1)}, &
 \Omega_{\text{GW}}(\eta, k) = & Q(n_{\text{s}}) A_\zeta^2 \left( \frac{k}{k_*} \right)^{2(n_{\text{s}}-1)},
\end{align}
where examples of the overall coefficient $Q(n_{\text{s}})$ are shown in Table~\ref{tab:overall_coefficient}.
For the central value  of the Planck 2015 TT$+$lowP constraint, $n_{\text{s}}=0.9655\pm 0.0062$~\cite{Ade:2015lrj}, and $Q(0.9655)= 0.8149$. %0.8149151673965718
 
Too large or small $n_{\text{s}}$ makes the integral divergent.
%%%%%%
\begin{table}[htb]
\begin{center}
\caption{The overall coefficient of the second-order GW sourced from the power-law index spectrum
}
  \begin{tabular}{|c||c|c|c|c|c|c|c|c|c|c|c|}
  \hline
$n_{\text{s}}$ & 0.4& 0.6& 0.8& 1.0 & 1.2 & 1.4 & 1.6 & 1.8 & 2.0 & 2.2  & 2.4\\
\hline
$Q(n_{\text{s}})$ & 0.8196 & 0.7984 & 0.7956 & 0.8222 & 0.8988 & 1.074 & 1.470 & 2.478 & 5.783 & 24.77 & 708.2 \\
\hline
  \end{tabular}
  \label{tab:overall_coefficient}
  \end{center}
\end{table}
%%%%%

%%%%%%%%%%%%%%%%%%%%%%%%%%%%%%
\subsection{Matter-dominated Universe}\label{ssec:MD}

The GW spectrum from the curvature perturbations in a MD Universe was studied in Refs.~\cite{Baumann:2007zm, Assadullahi:2009nf}, and the analytic formula for $I(v,u,x)$ was obtained there.
We also derive the formula using our conventions for self-completeness, which makes comparisons with other papers easier, and obtain fully analytic formulas of the GW power spectrum for simple examples, some of which were obtained only approximately.

In the MD Universe, the solution of eq.~\eqref{EOM_Green} for the Green's function of GW is 
\begin{align}
kG_{\bf k}(\eta, \bar{\eta}) =& - x \bar{x} ( j_1 (x)y_1(\bar{x}) - y_1 (x) j_1 (\bar{x})) \nonumber \\
=& \frac{1}{x \bar{x}} \left( (1+x\bar{x}) \sin (x- \bar{x}) - (x-\bar{x}) \cos (x-\bar{x}) \right).
\end{align}
For a late-time $\eta \gg \bar{\eta}$ and for a sufficiently large $k$, it is almost the sinusoidal functions, $-\frac{1}{\bar{x}}\cos (x-\bar{x})$ and $\sin (x - \bar{x})$, respectively.
The solution of eq.~\eqref{EOM_Phi} for the gravitational potential which is regular at $x \to 0$ is\footnote{This heuristic derivation is actually not a proper treatment because small perturbations to the pure MD equation affect properties of its solution significantly.  A proper treatment without neglecting the entropy perturbation shows the existence of the constant solution sourced by the entropy perturbation, which is a well-known fact in cosmology.} 
\begin{align}
\Phi(x) = 1.
\end{align}
The source function $f$ is
\begin{align}
f_{\text{MD}}(v,u,x) = \frac{6}{5}.
\end{align}
Since this is constant, the function $I(v,u,x)$ in the MD era can be much more easily obtained.
The ratio $a(\bar{\eta})/a(\eta)$ is now $(\bar{x}/x)^2$ in the MD era.  The function $I(v,u,x)$ turns out to be
\begin{align}
I_{\text{MD}}(v,u,x) = \frac{6(x^3+3x \cos x - 3 \sin x)}{5x^3}. \label{I_MD}
\end{align}
This asymptotes to $6/5$ in the large $x$ limit. For a small $x$, the leading term is $3 x^2 /25$.
When we introduce the oscillation average in eq.~\eqref{rho_GW}, we neglect the kinetic term and instead multiply the gradient term by 2.
To compensate for the factor 2 for the oscillation average of the nonoscillating term, we have to multiply the correction factor by $1/2$ to obtain
\begin{align}
\overline{I^2 (v,u,x\to \infty)} =& \frac{18}{25}.  \label{I_MD_osc_ave}
\end{align}

\paragraph{Example 1: Monochromatic source}
The first example for the curvature perturbation is the monochromatic case,
\begin{align}
\mathcal{P}_\zeta (k) = A_\zeta \delta \left( \log (k/k_*) \right).
\end{align}
The GW spectrum is obtained as 
\begin{align}
\Omega_{\text{GW}}=\frac{3}{25} \left( \frac{k_*}{a H} \right)^2 \left( 1- \left( \frac{k}{2k_*} \right)^2 \right)^2 A_\zeta ^2 \Theta (2k_* -k).
\end{align}

\paragraph{Example 2: Scale-invariant case}
If the MD era continues eternally, the density perturbations eventually become nonlinear.
Then the perturbation approach becomes invalid, so we set a cutoff scale $k_{\text{max}}$ to the curvature perturbations. Actually, the integral is divergent in the pure MD era unless we introduce such a cutoff.
In practice, the cutoff scale is the larger of the nonlinear scale $k_{\text{NL}}^{-1}\sim \mathcal{P}_\zeta ^{1/4} \mathcal{H}^{-1}$~\cite{Assadullahi:2009nf} (see also Appendix~\ref{sec:comparison}) and the scale corresponding to the onset of the MD era $k_{\text{MD}}^{-1}$ (for example, the beginning of the inflaton oscillation).
Thus, as a toy model, we assume a scale-invariant curvature perturbation with a cutoff $k_{\text{max}}$~\cite{Assadullahi:2009nf},
\begin{align}
\mathcal{P}_\zeta (k) = A_\zeta \Theta ( k_{\text{max}}- k).  \label{P_zeta_SI_cutoff}
\end{align}

\begin{figure}[tbh]
 \centering
{\includegraphics[width=0.65\columnwidth]{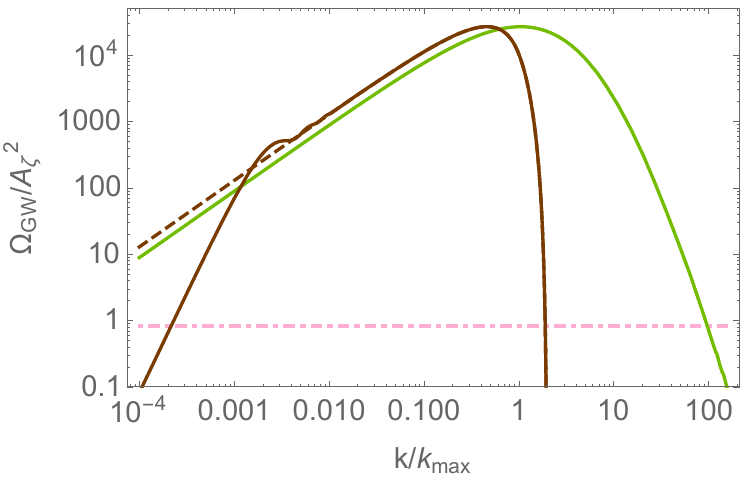}}
  \caption{The energy density fraction $\Omega_{\text{GW}}$ of GWs produced in the MD era from the scale-invariant source.
  \textbf{\textcolor{brown}{The brown lines}} (vanishing at $k=2 k_{\text{max}}$) show the case of an abrupt cutoff $k_{\text{max}}$~\eqref{P_zeta_SI_cutoff}, which may be interpreted as the scale corresponding to the beginning of inflaton oscillation or the scale where density perturbations become nonlinear.  The dashed curve represents eq.~\eqref{Omega_GW_MD_SI} where reheating is not considered (the pure MD case). The effect of reheating (transition to the RD era) is included in the solid line by multiplying eq.~\eqref{reheating_correction} for the case $k_{\text{R}}= 10^{-3} k_{\text{max}}$\protect \footnotemark.  These lines overlap for $k \gg k_{\text{R}}$.
  The above plot shows the spectra observed at late time ($\eta \gg \eta_{\text{R}}$) scaled back in time by the common redshift factor (independent of $k$) in such a way that they coincide with the spectra at $\mathcal{H}=  k_{\text{R}}$ for modes $k \gg k_{\text{R}}$. In other words, we have taken into account the evolution of modes $k \ll k_{\text{R}}$ since reheating to their horizon entry. 
  \textbf{\textcolor{wakakusa}{The green line}} shows the case of a MD era preceded by a RD era. $k_{\text{max}}$ is identified as $k_{\text{eq}}$. 
  The standard radiation-matter transition corresponds to this case with $k_{\text{eq}}=1.0\times 10^{-2} \, \text{Mpc}^{-1}$~\cite{Ade:2015xua}.  This line is obtained numerically using the interpolating transfer function~\eqref{Phi_interpolation}.
 Note that $\Omega_{\text{GW}}$ grows during the MD era.  The above plot shows the spectra at $\mathcal{H}= 10^{-3} k_{\text{max}}$ (see footnote~\ref{fn:nonlinearity}). 
  \textbf{\textcolor{sakura}{The pink line}} (horizontal dotted dashed) is the standard in the RD era [see eq.~\eqref{Omega_GW_RD_SI}] shown for comparison. 
  }
 \label{fig:Omega_GW_MD_SI}
  \end{figure}

For $0< k \leq k_{\text{max}}$, the integration regions dictated by the Heaviside theta function are 
$0 < t < \frac{2 k_{\text{max}}}{k} -2$, $-1 < s < 1$ and $\frac{2 k_{\text{max}}}{k} -2 < t < \frac{2 k_{\text{max}}}{k} -1$, $t- \frac{2 k_{\text{max}}}{k} +1 < s < -(t -\frac{2 k_{\text{max}}}{k} +1)$, 
while for $k_{\text{max}}<k\leq 2 k_{\text{max}}$, the integration region is
$0< t < \frac{2 k_{\text{max}}}{k} -1$, $t-\frac{2 k_{\text{max}}}{k} +1 < s < -(t-\frac{2 k_{\text{max}}}{k} +1)$. 
The GW strength is
\begin{align}
\Omega_{\text{GW}} = \frac{A_\zeta^2}{14000} \left(\frac{k}{aH} \right)^2  \times \begin{cases}
\left(1792 \tilde{k}^{-1} -2520 +768 \tilde{k} + 105 \tilde{k}^2 \right)   &   (0< k \leq k_{\text{max}})  \\
\left(1-2\tilde{k}^{-1} \right)^4  \left( 105 \tilde{k}^2 + 72  \tilde{k} + 16 -32 \tilde{k}^{-1}-16 \tilde{k}^{-2} \right)  &   (k_{\text{max}}<k\leq 2 k_{\text{max}})
\end{cases}, \label{Omega_GW_MD_SI}
\end{align}
where $\tilde{k} \equiv k/ k_{\text{max}}$.
The two expressions coincide up to and including the third derivative at $k=k_{\text{max}}$.
Equation~\eqref{Omega_GW_MD_SI} is shown as the dashed brown line in Figure~\ref{fig:Omega_GW_MD_SI}.

Note that the power spectrum $\mathcal{P}_h$ is enhanced by $\tilde{k}^{-1} = k_{\text{max}}/k$ for a small $k$.  This enhancement is due to the effect of a nondecaying scalar source, $\Phi=$const., during the MD era.  Taking the leading term for a small $k$ reproduces the result in Ref.~\cite{Assadullahi:2009nf} up to a numerical factor.\footnote{
After taking into account the difference of the normalization conventions, a factor $2$ is missed in the source side of their equation of motion for GW, and a geometric factor $\cos 2 \phi /\sqrt{2}$ is missed for projection to the transverse traceless mode where $\phi$ is the angle between the polarization vector $e({\bf k})$ and the projection of the source wavenumber $\tilde{k}$ onto the plane spanned by $e({\bf k})$ and $\bar{e}({\bf k})$.}

  \footnotetext{\label{fn:nonlinearity}
  For this choice with $A_\zeta \sim 10^{-9}$, the nonlinear scale $k_{\text{NL}}$ comes below $k_{\text{max}}$.  The reason for this choice is to clearly show the characteristic behavior of each of the modes $k \ll k_{\text{R}}$, $k_{\text{R}} \ll k < k_{\text{max}}$, and $k_{\text{max}}<k$.
  }

%%%%%%%%%%%%%%%%%%%%%%%%%%%%%%
\subsection{Transitions between radiation/matter eras}\label{ssec:general}

So far, we have considered the pure RD and the pure MD Universe.
However, the RD epoch is taken over by the late-time MD epoch. 
Also, an early MD era such as an inflaton oscillation period may precede the RD era.
We consider these transitions, the MD era to the RD era and the RD era to the MD era, separately below.
When one considers nonminimal cosmological scenarios, there may be multiple transitions. Generalization to such cases is a straightforward task.

\subsubsection{MD-to-RD transition}
We imagine a MD era dominated by some massive field which decays to reheat the Universe.
After the decay, it is a RD era.
We indicate the reheating time by the subscript R.
Before reheating ($x<x_{\text{R}} \equiv k \eta_{\text{R}}$), the function $I(v,u,x)$ is the same as the MD case we have seen above, \textit{i.e.}~$I(v,u,x)=I_{\text{MD}}(v,u,x)$.
After reheating ($x>x_{\text{R}}$), we separate the time integral as follows:
\begin{align}
I(v,u,x) =&\int_0^{x_{\text{R}}} \text{d}\bar{x} \left( \frac{x_{\text{R}}}{x} \right) \left( \frac{\bar{x}}{x_{\text{R}}} \right)^2  k G_k ^{\text{MD} \to \text{RD}}(\eta, \bar{\eta}) f_{\text{MD}}(v,u,\bar{x})   \, \nonumber \\
& + \int_{x_{\text{R}}}^x \text{d}\bar{x} \left( \frac{\bar{x}}{x} \right) k G_k ^{\text{RD}}(\eta, \bar{\eta}) f_{\text{MD} \to \text{RD}}(v,u,\bar{x})  , \label{I_MD2RD}
\end{align}
where the first line is the contribution from the MD era taking into account the fact that the propagation of the GW changes after reheating. The second line is the contribution from the RD era taking into account the fact that the scalar source experienced the MD era.

We may integrate it explicitly, but the expression is complicated.  Here, let us focus on terms with a qualitatively new behavior, that is, a feature beyond the simple sum of the RD and MD contributions.
Such a nontrivial feature resides in modes which were about to grow near the end of the MD era. 
These come from the first line of eq.~\eqref{I_MD2RD}.
We connect the GW solution at the transition requiring continuity of the zeroth and first derivatives.
Then the first line becomes
\begin{align}
&\int_0^{x_{\text{R}}} \text{d}\bar{x} \left( \frac{x_{\text{R}}}{x} \right) \left( \frac{\bar{x}}{x_{\text{R}}} \right)^2  k G_k ^{\text{MD} \to \text{RD}}(\eta, \bar{\eta}) f_{\text{MD}}(v,u,\bar{x}) \nonumber \\
&=\frac{3}{5 x x_{\text{R}}^3}\left( 3(2x_{\text{R}}^2-1)\cos x -6x_{\text{R}}\sin x +2x_{\text{R}}^4 \cos (x-x_{\text{R}})+ 4 x_{\text{R}}^3 \sin (x-x_{\text{R}})+ 3 \cos (x-2x_{\text{R}}) \right).  \label{MD2RD_nontrivial_piece}
\end{align}
In the limit $x_{\text{R}} \to x$, this reduces to the pure MD case result, eq.~\eqref{I_MD}.
On the other hand, in the limit $x \gg x_{\text{R}}$, it is approximated as $(12/25)(x_{\text{R}}^2/x) \sin x$.
By taking the square and oscillation average for $x$ and dividing it by eq.~\eqref{I_MD_osc_ave} and the common redshift factor $(x_{\text{R}}/x)$, we obtain the relative factor $R$ explaining the inefficient enhancement of superhorizon modes at reheating,
\begin{align}
R=& \frac{1}{4 x_{\text{R}}^8} \left(  \left( -6 x_{\text{R}} + 4x_{\text{R}}^3 \cos x_{\text{R}} + 2 x_{\text{R}}^4 \sin x_{\text{R}} + 6 \cos x_{\text{R}} \sin x_{\text{R}} \right)^2 \right. \nonumber \\
& \left. \qquad + \left( -3+6x_{\text{R}}^2 + 2 x_{\text{R}}^4 \cos x_{\text{R}} + 3 \cos 2 x_{\text{R}} -4 x_{\text{R}}^3 \sin x_{\text{R}}  \right)^2  \right), \label{reheating_correction}
\end{align}
which reduces to one in the subhorizon limit $x_{\text{R}} \gg 1$ and is proportional to $x_{\text{R}}^2$ in the superhorizon limit $x_{\text{R}}\ll 1$.
Multiplying eq.~\eqref{Omega_GW_MD_SI} by this factor, we obtain the brown solid line in Figure~\ref{fig:Omega_GW_MD_SI}.

The contribution to $I(v,u,x)$ from modes entering the horizon a bit after reheating scales like $k$. Squaring this and multiplying the integration region of $t$ which scales like $k_{\text{max}}/k$, the power spectrum for this range of wavenumber behaves like $\mathcal{P}_h \sim  (a(\eta_{\text{R}})/a(\eta))^2 (k_{\text{max}} k / k_{\text{R}}^2 )\mathcal{P}_\zeta^2(k)$, where we have replaced $\eta_{\text{R}} \sim k_{\text{R}}^{-1}$ with $k_{\text{R}} \equiv \mathcal{H} (\eta_{\text{R}})$.  This scaling is valid at late times $\eta \gg \eta_{\text{R}}$ since we have taken the late-time limit.  Note that the snapshot of the power spectrum at the time of reheating scales as $k^3$, but the observed spectrum scales as $k$, taking into account the evolution of modes since reheating to their horizon entry.  The origin of this evolution is the kinetic energy of GWs already developed at reheating. It seems that this change of scaling has not been explicitly noticed in the literature. 

 For shorter length scales, $ k_{\text{R}}< k < k_{\text{max}}$, it is similar to the MD case, $\mathcal{P}_h \sim (a(\eta_{\text{R}})/a(\eta))^2 (k_{\text{max}}/k) \mathcal{P}_\zeta^2(k)$.  On the other hand, for larger length scales, $k < k_{\text{its}}^{\text{R}}$ with the intersection $k_{\text{its}}^{\text{R}} \equiv k_{\text{R}} (k_{\text{R}}/k_{\text{max}})^{1/3}$, it is similar to the RD case, $\mathcal{P}_h \sim (a(\eta_{\text{R}})/a(\eta))^2 (k_{\text{R}}/k)^2 \mathcal{P}_\zeta^2(k) $.
The common factor $(a(\eta_{\text{R}})/a(\eta))^2$ represents redshift, and $\mathcal{P}_\zeta^2 (k)$ represents the source characteristics. The remaining factor represents the specific feature for the MD, RD, or intermediate era.

\subsubsection{RD-to-MD transition}
We now consider a RD era followed by a MD era denoting the equality time by the subscript eq.
Before the equality (\textit{i.e.}~$x < x_{\text{eq}} \equiv k \eta_{\text{eq}}$),  $I(v,u,x)=I_{\text{RD}}(v,u,x)$ is satisfied. After the equality, we may split the time integral for the function $I(v,u,x)$ as follows:
\begin{align}
I(v,u,x) =&\int_0^{x_{\text{eq}}} \text{d}\bar{x} \left( \frac{x_{\text{eq}}}{x} \right)^2 \left( \frac{\bar{x}}{x_{\text{eq}}} \right)  k G_k ^{\text{RD} \to \text{MD}}(\eta, \bar{\eta}) f_{\text{RD}}(v,u,\bar{x})  \, \nonumber \\
& + \int_{x_{\text{eq}}}^x \text{d}\bar{x} \left( \frac{\bar{x}}{x} \right)^2 k G_k ^{\text{MD}}(\eta, \bar{\eta}) f_{\text{RD} \to \text{MD}}(v,u,\bar{x})  . \label{I_RD2MD}
\end{align}
The first line is the contribution produced in the RD era when taking into account the fact that the GW propagation changes after the radiation-matter equality.
The second line is the contribution produced in the MD era when taking into account the fact that the source term experienced the RD era.

Again, for modes entering the horizon well before equality and well after equality, the power spectrum is essentially the same as that in the RD era and in the MD era, respectively.
This time, the only nontrivial terms come from the second line.
The scalar modes entering the horizon a bit before equality are suppressed in the RD Universe, so, even after the enhancement in the MD era, the corresponding GW power spectrum is less enhanced compared to the modes entering the horizon after equality.
Note also that this effect gives a physical cutoff for an otherwise divergent integral in the MD era.
Quantitatively, this effect is explained by the large $k$ limit of the transfer function,
\begin{align}
\Phi (\eta \gg \eta_{\text{eq}}, k \gg k_{\text{eq}}) = & \frac{\ln( c_1 k \eta_{\text{eq}})}{(c_2 k\eta_{\text{eq}})^2},  &
c_1 =& \frac{2}{\sqrt{3}(\sqrt{2}-1)} e^{\gamma - \frac{7}{2}} \approx 0.15,  & 
c_2 = & \frac{\sqrt{9/10}}{9 (\sqrt{2}-1)} \approx 0.25. \label{Phi_RD2MD_largek}
\end{align}
This is a standard result: see \textit{e.g.}~Ref.~\cite{Mukhanov:2005sc}.
Using this, the second line of eq.~\eqref{I_RD2MD} becomes
\begin{align}
&  \int_{x_{\text{eq}}}^x \text{d}\bar{x} \left( \frac{\bar{x}}{x} \right)^2 k G_k ^{\text{MD}}(\eta, \bar{\eta}) f_{\text{RD} \to \text{MD}}(v,u,\bar{x}) \nonumber \\
&=\frac{(x^3-3(x-x_{\text{eq}})-x x_{\text{eq}}^2)\cos(x-x_{\text{eq}}) -(3+3x x_{\text{eq}}-x_{\text{eq}}^2)\sin (x-x_{\text{eq}}) }{x^3} \times \frac{6}{5} \frac{\ln( c_1 u x_{\text{eq}})}{(c_2 u x_{\text{eq}})^2} \frac{\ln( c_1 v  x_{\text{eq}})}{(c_2 v x_{\text{eq}})^2} ,   \label{RD2MD_nontrivial_piece}
\end{align}
For a late time $\eta \gg \eta_{\text{eq}}$, the first factor asymptotes to 1, and the $k$ dependence at large $k$ is given just by $\Phi^2$, or $k^{-4} (\ln k)^2$.  
Squaring this and numerically integrating\footnote{
For this purpose, we interpolate the large $k$ behavior of the transfer function~\eqref{Phi_RD2MD_largek} and the small $k$ limit $\Phi = 1$ with the following function:
\begin{align}
\Phi (\eta \gg \eta_{\text{eq}} , k) =  \frac{\log (1 + c_1 k \eta_{\text{eq}})}{\log (1 + c_1 k \eta_{\text{eq}}) +(c_2 k \eta_{\text{eq}})^2 }. \label{Phi_interpolation}
\end{align}
The numerical result of $\Omega_{\text{GW}}$ using this transfer function in the case of the scale-invariant source is shown as the green solid line in Figure~\ref{fig:Omega_GW_MD_SI}.  
}
%%%
 it with respect to $t$ and $s$, we find that the power spectrum scales as $\mathcal{P}_h \sim (k\eta_{\text{eq}})^{-2 \gamma(k)}  \mathcal{P}_\zeta^2 (k) $ with $3 \lesssim \gamma (k) \lesssim 4 $ being an increasing function of $k$.
This is consistent with an observation in Ref.~\cite{Baumann:2007zm}.
For larger length scales, $k< k_{\text{eq}}$, it is essentially the MD era, and $\mathcal{P}_h \sim (k_{\text{eq}}/k) \mathcal{P}_\zeta^2(k)$, where $k_{\text{max}}$ has been replaced by the physical cutoff $k_{\text{eq}}$.
For shorter length scales, $k_{\text{eq}} < k < k_{\text{its}}^{\text{eq}}$ with the intersection $k_{\text{its}}^{\text{eq}} \equiv k_{\text{eq}} (k_{\text{eq}}/\mathcal{H}(\eta))^{2/(\gamma -1)} $, it is essentially the RD era, so $\mathcal{P}_h \sim (a(\eta_{\text{eq}})/a(\eta))^2 (k_{\text{eq}}/k)^2 \mathcal{P}_\zeta^2 (k)$.

%%%%%%%%%%%%%%%%%%%%%%%%%%%%%%%%%%%%%%%%%%%%%%%%%%%
\section{Summary and conclusion}\label{sec:conclusion}
Traditionally, the second-order GWs sourced from the primordial curvature perturbations have been studied mainly numerically.
We have analytically calculated the part of the curvature-induced GW power spectrum $I(v,u,x)$ [defined in eq.~\eqref{I}], which is calculable independently of the primordial curvature perturbations $\mathcal{P}_\zeta (k)$.  One of our main results is the expression~\eqref{I_RD} and its late-time oscillation average, $\overline{I^2_{\text{RD}}(v,u,x)}$  [eq.~\eqref{I_RD_osc_ave}], or, equivalently in terms of the other variables, $\overline{I^2_{\text{RD}}(t,s,x)}$ [eq.~\eqref{I_RD_osc_ave_ts}].
Once the primordial curvature perturbation $\mathcal{P}_\zeta (k)$ is specified, one can easily compute the remaining integrals for $u$ and $v$ [eq.~\eqref{P_h}], or, equivalently, $t$ and $s$ [eq.~\eqref{P_h_ts}] whose physical meaning is the wavenumber of the gravitational potential $\Phi$.

As applications of the analytic formula, we have calculated the power spectrum of the induced GW for simple examples of the primordial curvature perturbations. This has been done numerically or fully analytically when possible.
For completeness, we have also studied the second-order GWs induced in a MD era and have obtained analytic formulas for simple examples.
Moreover, we have suggested an approximate way of analytically calculating the GW power spectrum in the presence of transitions between the RD and MD eras.
In fact, using our formulas, we have derived the nontrivial wavenumber dependence of the induced GW power spectrum.
In particular, we have analytically obtained the suppression factor~\eqref{reheating_correction} for modes entering the horizon after reheating by assuming the sudden transition between the MD and RD eras, taking into account the growth of these modes after reheating until their horizon entry. 
The RD-to-MD transition can be treated numerically.
In this way, the nontrivial shape of the power spectrum of the induced GWs in the presence of finite duration of the MD era, during which the GW spectrum is enhanced, is obtained (semi)analytically, as demonstrated in Figure~\ref{fig:Omega_GW_MD_SI} for the case of the scale-invariant curvature perturbations.

Our results are useful when one quantitatively evaluates the power spectrum $\mathcal{P}_h (\eta, k)$ or the corresponding energy fraction parameter $\Omega_{\text{GW}}(\eta, k)$ of the GW induced from the curvature perturbations. 
These quantities are to be compared with observations as illustrated in Appendix~\ref{sec:comparison} and Figure~\ref{fig:Illustration} with simple examples.  
From Figure~\ref{fig:Illustration}, one can see that it would be difficult to detect the induced GWs by near future observations if the curvature perturbation can be approximated as a scale-invariant one.
This is so even if we assume the presence of an early MD era to enhance the induced GWs, as long as we consider the linear regime.
By contrast, if the spectrum of the curvature perturbation has a sufficiently blue tilt or running, it may be possible to detect it as shown in the Figure~\ref{fig:Illustration}.
In such a case, one has to consider constraints on the enhanced curvature perturbations by $\mu$ distortion of the CMB~\cite{Chluba:2012we, Kohri:2014lza}, change of the baryon-to-photon ratio~\cite{Nakama:2014vla} and the neutron-to-proton ratio~\cite{Inomata:2016uip} (see also Ref.~\cite{Jeong:2014gna}) in BBN, and overproduction of ultracompact minihalos~\cite{Lacki:2010zf, Bringmann:2011ut} or PBHs (for reviews of the constraints, see \textit{e.g.}~Refs.~\cite{Carr:2009jm, Carr:2016drx, Sasaki:2018dmp}).

%%%%%%%%%%%%%%%%%%%%%%%%%%%%%%%%%%%
\section*{Note added}

Ref.~\cite{Espinosa:2018eve} appeared recently after the most parts of
this paper had been completed.
They derived an analytic formula of the late-time limit of the
gravitational field $h_{\bf k}(\eta)$ induced in the radiation
dominated era right after inflation, from which the power spectrum as
well as bispectrum can also be calculated, in the context of the
Standard Model Higgs instability.

%%%%%%%%%%%%%%%%%%%%%%%%%%%%%%%%%%
\section*{Acknowledgments}
The authors are grateful to Sachiko Kuroyanagi for reading the manuscript and for the useful discussion.
TT thanks Kyohei Mukaida for explaining the conventions in Ref.~\cite{Inomata:2016rbd}.
This work is supported in part by the JSPS Research Fellowship for Young
Scientists (TT) and JSPS KAKENHI Grants No.~JP17J00731 (TT),
No.~JP17H01131 (KK), and No.~26247042 (KK), and MEXT KAKENHI Grants No.~JP15H05889
(KK), No.~JP16H0877 (KK), and No.~JP18H04594 (KK).
%%%%%%%%%%%%%%%%%%%%%%%%%%%%%%%%%%%%%%%%%%%%%%%%%%%
\appendix
\section{General integral formulas with radiation/matter transitions}\label{sec:general}
In this Appendix, we provide general formulas which can be used for the calculation of $I(v,u,x)$ in the presence of multiple RD/MD transitions.
We need to consider a generalization of eqs.~\eqref{I_MD2RD} and \eqref{I_RD2MD}.

In general, in a RD era, $\Phi$ can be written as a linear combination of two independent solutions, $3j_1(x/\sqrt{3})/(x/\sqrt{3})$ and $3y_1 (x/\sqrt{3})/(x/\sqrt{3})$.
The GW solution is a linear combination of two independent solutions, $\sin \bar{x}$ and $\cos \bar{x}$.
On the other hand, in a MD era, $\Phi$ can be written as a linear combination of two independent solutions, $1$ and $x^{-5}$. (However, the power of the decaying mode changes when we perturb the pure MD case, and in any case, we neglect the decaying mode)
The GW solution is a linear combination of two independent solutions, $\bar{x}j_1 (\bar{x})$ and $\bar{x}y_1 (\bar{x})$.

For the RD case, we consider
\begin{align}
\mathcal{I}_{\text{RD}}(v,u,x_1,x_2)=\int_{x_1}^{x_2} \text{d}\bar{x} \, \bar{x}  (C \sin \bar{x} + D \cos \bar{x} ) f_{\text{RD}}(v,u,\bar{x})|_{\Phi(\bar{x}) = 3\sqrt{3}(A j_1 (\bar{x}/\sqrt{3})+B y_1 (\bar{x}/\sqrt{3}))/\bar{x}}.
\end{align}
For the MD case, we consider
\begin{align}
\mathcal{I}_{\text{MD}}(v,u,x_1,x_2)=\int_{x_1}^{x_2} \text{d}\bar{x} \, \bar{x}^2  (C \bar{x} j_1 (\bar{x}) + D \bar{x} y_1(\bar{x}) ) f_{\text{MD}}(v,u,\bar{x})|_{\Phi(\bar{x}) = A 
}, 
\end{align}
where $x_1 \equiv k \eta_1$ and $x_2 \equiv k \eta_2$, and where $A, B, C$, and $D$ are constants with respect to $\bar{\eta}$.
$A$ and $B$ may depend on $k$ and $\eta_1$, and $C$ and $D$ may depend on $k$, $\eta_2$, and the present time $\eta$.
Substituting $\Phi$ into $f(v,u,k,\eta)$, $A(k)$ becomes $A(u k)$ or $A(v k)$, and $B$ behaves similarly.
We show only the dependence on $u$ and $v$, so we express this as $A(u)$ and $A(v)$.

The function $f$ is now
\begin{align}
& f_{\text{RD}}(v,u,\bar{x})|_{\Phi(\bar{x}) = 3\sqrt{3}(A j_1 (\bar{x}/\sqrt{3})+B y_1 (\bar{x}/\sqrt{3}))/\bar{x}} \nonumber \\
&= \frac{1}{u^3 v^3 \bar{x}^6} \left( E^{\cos,v-u} \cos \frac{(v-u)\bar{x}}{\sqrt{3}} +E^{\sin,v-u} \sin \frac{(v-u)\bar{x}}{\sqrt{3}}+E^{\cos,v+u} \cos \frac{(v+u)\bar{x}}{\sqrt{3}}+E^{\sin,v+u} \sin \frac{(v+u)\bar{x}}{\sqrt{3}}  \right),
\end{align}
where the $E$ values are functions of $(u,v,\bar{x})$,
\begin{align}
E^{\cos,v-u} =& 6\left(A(u)A(v)+B(u)B(v)\right) \left(  54- 6 (u^2 + v^2 - 3 u v) \bar{x}^2 + u^2 v^2 \bar{x}^4\right) \nonumber \\
& \qquad + 12 \sqrt{3} \left(A(u)B(v)-A(v)B(u) \right)(u-v)\bar{x}(9+u v \bar{x}^2) , \\
E^{\cos,v+u} =& -6 \left( 2\sqrt{3} \left(A(u)B(v)+A(v)B(u)\right) (u+v) \bar{x} (-9 +uv \bar{x}^2 ) \right. \nonumber \\
& \left. \qquad + \left(A(u)A(v)-B(u)B(v)\right) (54-6(u^2+v^2+3uv)\bar{x}^2+u^2 v^2 \bar{x}^4) \right) , \\
E^{\sin,v-u} =& -6 \left( 2 \sqrt{3} \bar{x} \left(A(u)A(v) + B(u)B(v)\right) (u-v) (9+uv\bar{x}^2) \right. \nonumber \\
& \left.  \qquad + \left( A(v)B(u) - A(u) B(v)  \right) \left( 54-6(u^2-3u v +v^2 )\bar{x}^2 + u^2 v^2 \bar{x}^4 \right) \right), \\
E^{\sin, v+u}=&6 \left(2 \sqrt{3}\left(A(u)A(v) -B(u)B(v)\right)  (u+v) \bar{x}(-9+uv\bar{x}^2) \right. \nonumber \\
& \left. \qquad + \left(A(u)B(v)+A(v)B(u)\right)(-54+6(u^2+v^2+3 uv)\bar{x}^2 -u^2 v^2 \bar{x}^4) \right).
\end{align}
The counterpart in the MD case is
\begin{align}
 f_{\text{MD}}(v,u,\bar{x})|_{\Phi(\bar{x}) =A
 }=& \frac{6A(u)A(v)}{5}  .
  \end{align}

The integrals $\mathcal{I}(v,u,x_1, x_2)$ are as follows:
\begin{align}
\mathcal{I}_{\text{RD}}(v,u,x_1,x_2)=& \frac{3}{4u^3 v^3} \left [ \frac{1}{x^4} \left( F^{--} \cos y^{--}  +F^{+-} \cos y^{+-}  +F^{-+} \cos y^{-+}  +F^{++} \cos y^{++}   \right. \right. \nonumber \\ 
& \qquad \left. \phantom{\frac{1}{x^4}} \left. + G^{--} \sin y^{--} + G^{+-} \sin y^{+-} + G^{-+} \sin y^{-+} + G^{++} \sin y^{++} \right) \right]_{x_1}^{x_2}  \nonumber \\
& + \frac{3(u^2+v^2-3)^2}{4u^3 v^3} \left [ H^{--} \text{Ci}(y^{--}) + H^{+-}\text{Ci}(y^{+-}) +H^{-+}  \text{Ci}(|y^{-+}|)  +H^{++} \text{Ci}(y^{++}) \right. \nonumber \\
&\left. \qquad \qquad +I^{--} \text{Si}(y^{--}) +I^{+-} \text{Si}(y^{+-}) +I^{-+} \text{Si}(y^{-+})+ I^{++} \text{Si}(y^{++})  \right]_{x_1}^{x_2},
\end{align}
where we have introduced $y^{\pm\pm}=\left(1 \pm \frac{v\pm u}{\sqrt{3}}\right)x$ for compact notation, where the first (second) $\pm$ on the left side corresponds to the first (second) $\pm$ on the right side.  (The first sign is the relative sign between 1 and $v$, and the second sign is the relative sign between $v$ and $u$.)
The coefficient functions $F$, $G$, $H$, and $I$ are defined as
\begin{align}
F^{--}=& I^{--} \left( 18(-1+\sqrt{3}(u-v))x + (-3+\sqrt{3}(u-v))((u+v)^2-3)x^3 \right) \nonumber \\
&  - H^{--} \left( 54 - 3 (3+ u^2 + v^2 -6 u v + 2\sqrt{3} (v-u))x^2 \right) \\
F^{+-}=& -I^{+-} \left( 18(1+\sqrt{3}(u-v))x + (3+\sqrt{3}(u-v))((u+v)^2-3)x^3 \right) \nonumber \\
&  - H^{+-} \left( 54 - 3 (3+ u^2 + v^2 -6 u v + 2\sqrt{3} (u-v))x^2 \right) \\
F^{-+}=& -I^{-+} \left( 18(1+\sqrt{3}(u+v))x + (3+\sqrt{3}(u+v))((u-v)^2-3)x^3 \right) \nonumber \\
&  - H^{-+} \left( 54 - 3 (3+ u^2 + v^2 +6 u v + 2\sqrt{3} (u+v))x^2 \right) \\
F^{++}=& I^{++} \left( 18(-1+\sqrt{3}(u+v))x + (-3+\sqrt{3}(u+v))((u-v)^2-3)x^3 \right) \nonumber \\
&  - H^{++} \left( 54 - 3 (3+ u^2 + v^2 +6 u v - 2\sqrt{3} (u+v))x^2 \right) \\
G^{--}=& - H^{--} \left( 18(-1+\sqrt{3}(u-v))x + (-3+\sqrt{3}(u-v))((u+v)^2-3)x^3 \right) \nonumber \\
&  - I^{--} \left( 54 - 3 (3+ u^2 + v^2 -6 u v + 2\sqrt{3} (v-u))x^2 \right) \\
G^{+-}=& H^{+-} \left( 18(1+\sqrt{3}(u-v))x + (3+\sqrt{3}(u-v))((u+v)^2-3)x^3 \right) \nonumber \\
&  - I^{+-} \left( 54 - 3 (3+ u^2 + v^2 -6 u v + 2\sqrt{3} (u-v))x^2 \right) \\
G^{-+}=& H^{-+} \left( 18(1+\sqrt{3}(u+v))x + (3+\sqrt{3}(u+v))((u-v)^2-3)x^3 \right) \nonumber \\
&  - I^{-+} \left( 54 - 3 (3+ u^2 + v^2 + 6 u v + 2\sqrt{3} (u+v))x^2 \right) \\
G^{++}=& -H^{++} \left( 18(-1+\sqrt{3}(u+v))x + (-3+\sqrt{3}(u+v))((u-v)^2-3)x^3 \right) \nonumber \\
&  - I^{++} \left( 54 - 3 (3+ u^2 + v^2 + 6 u v - 2\sqrt{3} (u+v))x^2 \right) \\
H^{--}=&  (A(u)A(v)+B(u)B(v))D +(A(u)B(v)-A(v)B(u))C   , \\
H^{+-}=&  (A(u)A(v)+B(u)B(v)) D +(A(v)B(u)-A(u)B(v))C   , \\
H^{-+}=& - ((A(u)A(v)-B(u)B(v))D +(A(u)B(v)+A(v)B(u)) C) , \\
H^{++}=& -((A(u)A(v)-B(u)B(v))D-(A(u)B(v)+A(v)B(u))C)  , \\
I^{--}=& (A(u)A(v)+B(u)B(v))C +(A(v)B(u)-A(u)B(v))D  , \\
I^{+-}=& (A(u)A(v)+B(u)B(v))C +(A(u)B(v)-A(v)B(u))D  , \\
I^{-+}=& -((A(u)A(v)-B(u)B(v))C-(A(u)B(v)+A(v)B(u))D) , \\
I^{++}=& -((A(u)A(v)-B(u)B(v))C+(A(u)B(v)+A(v)B(u))D) .
\end{align}
The $x_1\to 0$ limit can be taken by using $\lim_{x_1\to 0} \text{Ci}(A x_1) - \text{Ci}(B x_1) = \log A - \log B$.
The above formula correctly reproduces earlier results.
For example, if we take $A=1$, $B=0$, $C=-\cos x$, and $D=\sin x$, then $x^{-1}\lim_{x_1 \to 0}\mathcal{I}_{\text{RD}}(v,u,x_1, x) = I_{\text{RD}}(v,u,x)$.

The MD counterpart of the integral is
\begin{align}
\mathcal{I}_{\text{MD}}(v,u,x_1,x_2)=& \frac{6A(u)A(v)}{5}  \left[ C \left( -3 x \cos x +(3-x^2)\sin x \right) +D\left(- 3 x \sin x + (x^2 -3) \cos x  \right) \right]^{x_2}_{x_1}  .
\end{align}
If we take $A=1$, $C=x y_1 (x)$, and $D=-x j_1 (x)$, this reduces to the pure MD result,\\ $x^{-2} \lim_{x_1 \to 0} \mathcal{I}_{\text{MD}} (v,u,x_1, x) = I_{\text{MD}}(v,u,x)$.
Equation~\eqref{MD2RD_nontrivial_piece} is obtained by using the values of $C$ and $D$ which equate the zeroth and first derivatives before and after the reheating (sudden decay approximation is used).
Also, we substitute $\Phi$ in eq.~\eqref{Phi_RD2MD_largek} to $A$ to obtain eq.~\eqref{RD2MD_nontrivial_piece}.

%%%%%%%%%%%%%%%%%%%%%%%%%%%%%%%%%%%%%%%%%%%%%
\section{Comparison with observations}
\label{sec:comparison}

\begin{figure}[tbh!]
 \centering
{\includegraphics[width=0.78 \columnwidth]{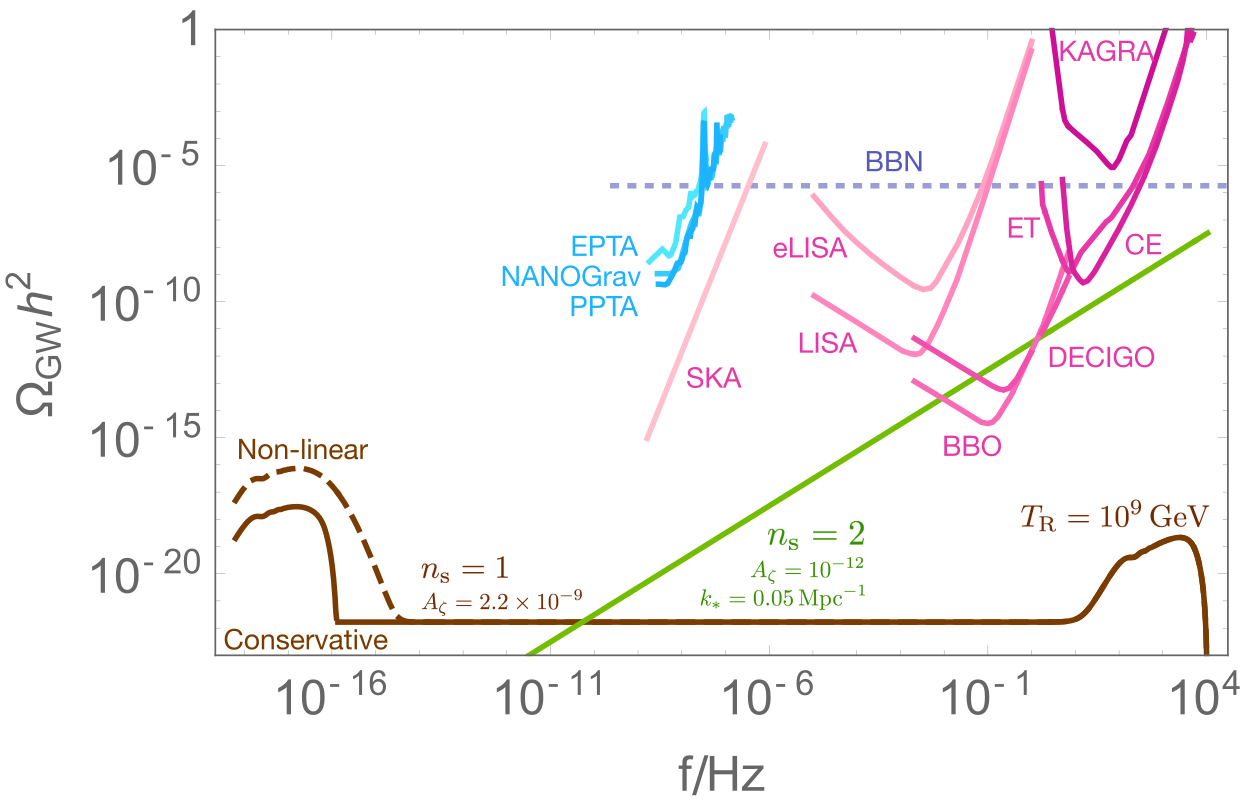}}
  \caption{Simple examples of the energy density fraction $\Omega_{\text{GW}} h^2$ of the induced GWs.
   \textbf{\textcolor{brown}{The brown lines}} show the case of the scale-invariant curvature perturbations with $A_\zeta = 2.2 \times 10^{-9}$. The horizontal part is the contribution from the RD era. The bottom left curves show the effect of the late-time MD era. The dashed line is in the nonlinear regime, and the solid line is a conservative one neglecting all of the contributions beyond the nonlinearity scale.  The bottom right curve shows an example of an early MD era with the reheating temperature $T_{\text{R}}= 10^9 \, \text{GeV}$.  The scale of the onset of the early MD era is assumed to be 200 times shorter than the reheating scale so that there is no nonlinearity issue.
   \textbf{\textcolor{wakakusa}{The green line}} shows the contribution from the RD era in the case of power-law curvature perturbations with $A_\zeta = 10^{-12}$, $n_{\text{s}}=2$, and $k_* = 0.05 \, \text{Mpc}^{-1}$. 
   \textbf{\textcolor{skyblue}{The blue lines}} denote existing pulsar timing array constraints from EPTA~\cite{Lentati:2015qwp}, NANOGrav~\cite{Arzoumanian:2018saf}, and PPTA~\cite{Shannon:2015ect}.
    \textbf{\textcolor{sakura}{The pink lines}}  show sensitivity curves~\cite{Sathyaprakash:2009xs} of various future GW observations reproduced from Ref.~\cite{Moore:2014lga}.  The observations are from SKA~\cite{5136190}, eLISA~\cite{Seoane:2013qna}, LISA~\cite{2017arXiv170200786A}, BBO~\cite{Harry:2006fi}, DECIGO~\cite{Seto:2001qf}, Einstein Telescope~\cite{Punturo:2010zz}, Cosmic Explorer~\cite{Evans:2016mbw}, and KAGRA~\cite{Somiya:2011np}. 
\textbf{\textcolor{purplegray}{The gray line}} (dotted) shows the upper bound on the relativistic degrees of freedom from BBN, $\Omega_{\text{GW}}h^2 < 1.8 \times 10^{-6}$ (95\% C.L.) derived in Appendix~\ref{sec:BBN}.
  }
 \label{fig:Illustration}
  \end{figure}

Although the focus of this paper is the derivation of the analytic formulas of the power spectrum of the induced GWs, we briefly illustrate how to compare our results to observations.  The related discussion is given at the end of section~\ref{sec:conclusion}.

Well after the horizon entry, GWs produced in a RD era redshift as radiation $\rho_{\text{GW}} \propto a^{-4}$, so $\Omega_{\text{GW}}$ is constant during a RD era, but it is diluted as $a^{-1}$ in a MD era. This fact is represented by the redshift factor $(x_{\text{eq}}/x)^2$ in eq.~\eqref{I_RD2MD}.
The present value of the energy fraction for the contribution from or before the RD era is thus
\begin{align}
\Omega_{\text{GW}}(\eta_0, k) = \Omega_{\text{r},0} \Omega_{\text{GW}} (\eta_{\text{c}}, k),
\end{align}
where $\Omega_{\text{r},0} = \rho_{\text{r},0}/\rho_0$ is the present value of the energy density fraction of radiation, and $\eta_{\text{c}}$ is some time after $\Omega_{\text{GW}}(\eta,k)$ has become constant so $\Omega_{\text{GW}} (\eta_{\text{c}}, k)$ is the asymptotic constant value during the RD era~\cite{Inomata:2017vxo}.
A precise formula taking into account the change in the number of relativistic degrees of freedom can be found, \textit{e.g.}, in Ref.~\cite{Jinno:2013xqa} in the context of the primordial GWs.
For a comprehensive discussion on the precise temperature dependence of the effective degrees of freedom, see Ref.~\cite{Saikawa:2018rcs}.
We do not show such a dependence here because we neglect such changes in the analytic integral; see footnote~\ref{fn:dof}.

On the other hand, 
the present value of the energy fraction for the contribution after the radiation-matter equality is obtained as
\begin{align}
\Omega_{\text{GW}}(\eta_0, k) = & \frac{\rho_{\text{GW}}(\eta_0, k)}{\rho_{\text{GW}}(\eta_\Lambda, k)} \frac{\rho_{\text{GW}}(\eta_\Lambda, k)}{\rho(\eta_\Lambda)} \frac{ \rho(\eta_\Lambda)}{\rho_m (\eta_0)} \frac{\rho_m (\eta_0)}{\rho (\eta_0)} \nonumber \\
\simeq & 2\, \Omega_{\text{m},0} \frac{a(\eta_\Lambda)}{a(\eta_0)} \Omega_{\text{GW}}(\eta_\Lambda, k),
\end{align} 
where $\Omega_{\text{m},0}=\rho_{\text{m},0}/\rho_0$ is the present value of the matter energy fraction, $\eta_{\Lambda}$ is the conformal time when the dark energy begins to dominate the Universe, $\rho_{\text{m}}(\eta_\Lambda)=\rho_{\Lambda} (\eta_\Lambda) \simeq \rho(\eta_\Lambda)/2$, and we approximate $\Omega_{\text{GW}}(\eta_\Lambda , k)$ by the MD-to-RD formula because GWs are supposed to redshift like radiation after the MD era. 

To suppress the uncertainty of the Hubble parameter, it is customary to multiply $\Omega_{\text{GW}}$ with $h_0^2$, which is defined as $H_0 = 100 \, h_0 \, \text{km}/\text{s} / \text{Mpc}$.
Some simple examples are plotted in Figure~\ref{fig:Illustration} for illustration.
The brown lines show the scale-invariant case with two MD eras.
The green line shows an example of the power-law spectrum, which may be interpreted qualitatively as a rough approximation for some PBH scenarios with curvatons~\cite{Kawasaki:2012wr, Kohri:2012yw, Carr:2017edp, Ando:2017veq}.  The pulsar timing array constraints and the sensitivity curves of future GW detectors are also shown, as blue and pink lines, respectively.

The dashed line in the Figure~\ref{fig:Illustration} indicates that it is in the nonlinear regime.
Since the gravitational potential and the density perturbation are related to each other through $\Delta \Phi \simeq a^2 \delta \rho /2$ in the deep subhorizon limit, we define the nonlinear scale as (see Ref.~\cite{Assadullahi:2009nf})
\begin{align}
k_{\text{NL}}(\eta) = \frac{3}{2} \mathcal{P}_\zeta^{-1/4} \mathcal{H}(\eta).
\end{align}
We are mostly interested in the nonlinear scale evaluated at the end of the MD era.
The solid line below the dashed one is the case in which we cut off the source spectrum at the nonlinear scale, and the line of the RD era is simply extrapolated.
This should be too conservative because there would be a contribution which gets marginally nonlinear during the MD era and subsequently diluted just by cosmic expansion.  This contribution scales as $(k_{\text{NL}}/k)^4$~\cite{Assadullahi:2009nf}.  However, to derive the precise spectrum, including the region around $k\simeq k_{\text{NL}}$, one has to consider a time-dependent cutoff $k_{\text{NL}}(\eta)$ or rely on nonlinear lattice simulations.
For simplicity, we neglect this contribution. The true value will be between the dashed and solid curves.

%%%%%%%%%%%%%%%%%%%%%%%%%%%%%%%%%%%%%%%%%%%%%%%%%%%%%%%%%%%%%%%%%%%%%%
\section{Constraints on GW  from Big-Bang Nucleosynthesis}
\label{sec:BBN}
%%%%%%%%%%%%%%%%%%%%%%%%%%%%%%%%%%%%%%%%%%%%%%%%%%%%%%%%%%%%%%%%%%%%%%

An extra component of radiation such as the primordial gravitational
wave background speeds up the expansion of the Universe, which can be
checked by light element abundances produced in the epoch of BBN. Such an extra component of radiation is often
parametrized by the effective number of neutrino species
$N_{\nu, \rm eff} \equiv \rho_{\nu, \rm eff}/\rho_{\nu_i}$, where
$\rho_{\nu, \rm eff}$ is the total energy density for the three
species of active neutrinos and the extra component of radiation, and
$\rho_{\nu_i}$ is the energy density for one species of active neutrino
$\nu_i$. If $N_{\nu, \rm eff}$ is larger than $\sim 3$, the
interconverting reactions between neutron ($n$) and proton ($p$)
should be decoupled from the thermal bath earlier than the time in the case of
the standard BBN, which gives a larger neutron to proton ratio ($n/p$)
as its freeze-out value. Then more $^4$He and D are produced due to
this larger $n/p$. Compared with observational light element
abundances of $^4$He and D, we can constrain $N_{\nu, \rm eff}$ for a
fixed value of baryon number. Here, we adopt the value of the baryon
number to be $\Omega_{\rm B}h_0^2 =0.02229 ^{+0.00029}_{-0.00027}$ (95$\%$
C.L.)~\cite{Ade:2015xua}.

\begin{figure}[tbh!]
 \centering
    \subcaptionbox{ D and $^4$He \label{sfig:chi2Nnu}}
{\includegraphics[width=0.48\columnwidth]{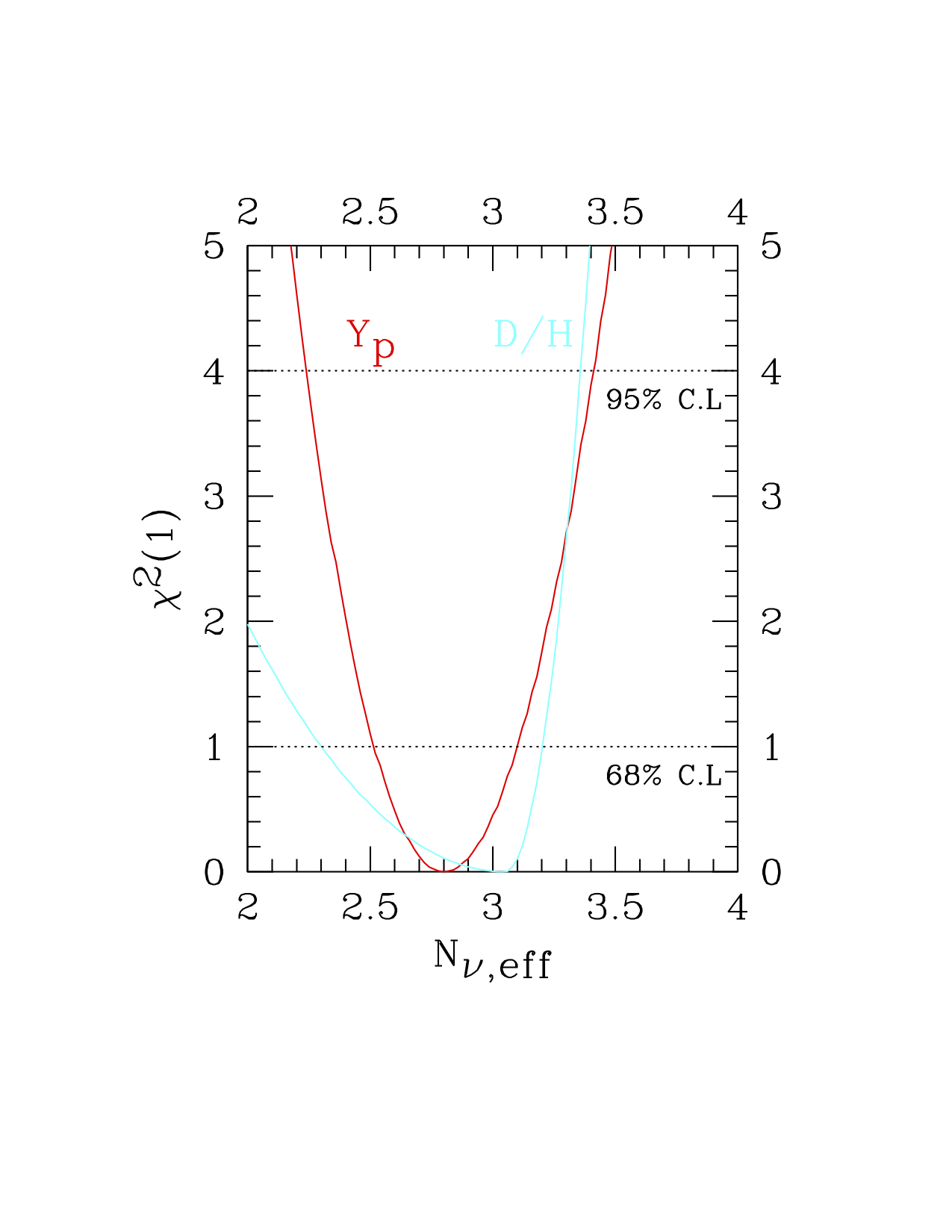}}~
    \subcaptionbox{ Total  \label{sfig:chi2OmegaGWv2}}
{\includegraphics[width=0.48\columnwidth]{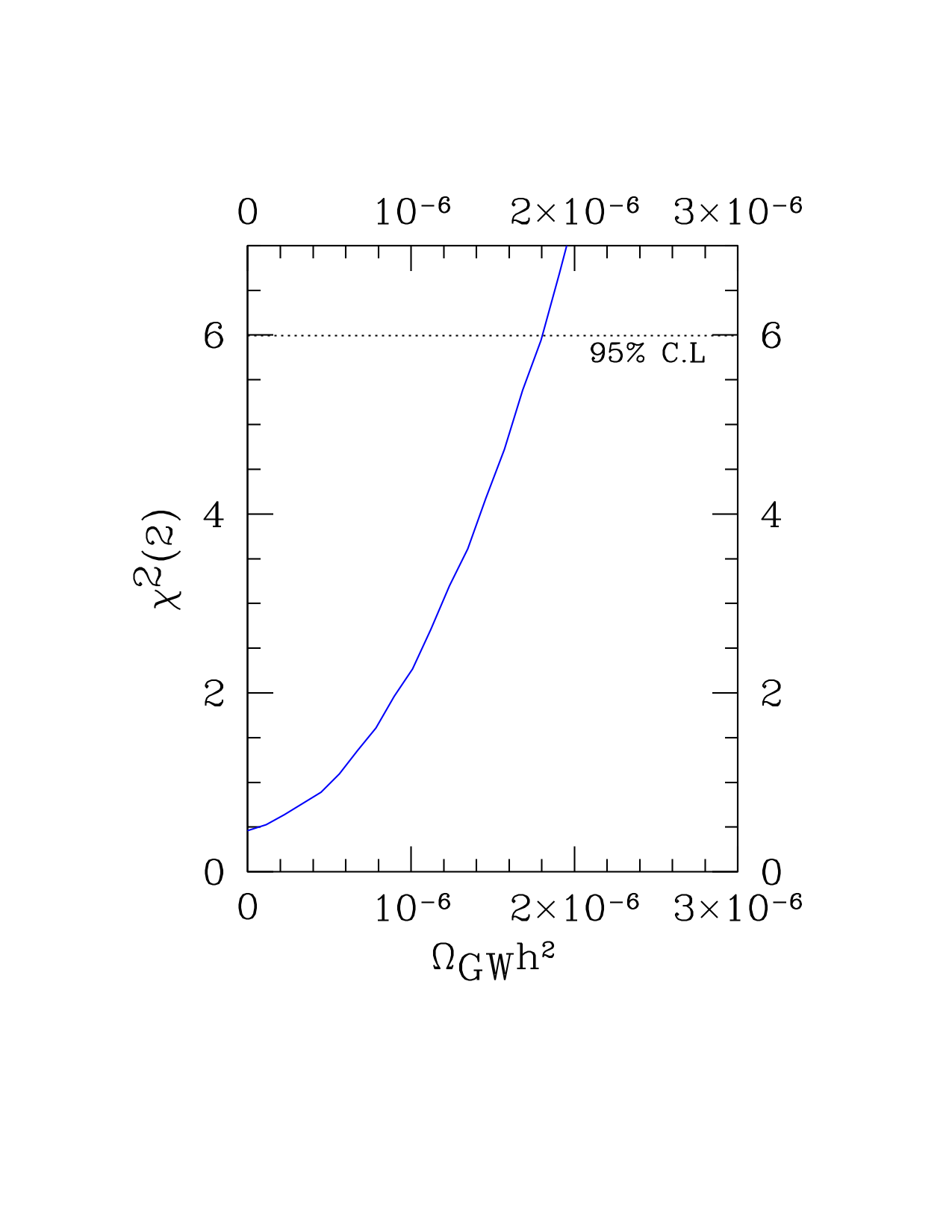}}
  \caption{\eqref{sfig:chi2Nnu} $\chi^2$ as a function of $N_{\nu, \rm eff}$ to fit abundances
  of D and $^4$He, respectively.  \eqref{sfig:chi2OmegaGWv2} Total $\chi^2$ as a function of $\Omega_{\rm GW}h_0^2$ to  simultaneously fit both D and $^4$He.
  }
 \label{fig:I}
  \end{figure}

In this paper we adopt the following observational values of the mass
fraction of $^4$He~\cite{Aver:2015iza} and the deuterium (D) to hydrogen
(H) ratio~\cite{Zavarygin:2018dbk} at 68$\%$ C.L.,
\begin{eqnarray}
  \label{eq:Yp}
  {\rm Y}_p = 0.2449 \pm 0.0040,
\end{eqnarray}
and
\begin{eqnarray}
  \label{eq:Deuterium}
  \left({\rm D/H} \right)_p = (2.545 \pm 0.025) \times 10^{ -5 },
\end{eqnarray}
respectively. 

In Fig.~\ref{sfig:chi2Nnu}, we plot $\chi^2$s as a function of
$N_{\nu, \rm eff}$ to fit the observational abundance of D and $^4$He,
respectively, by using theoretical values of abundances calculated in
BBN with errors of nuclear reaction rates. By using these values of $\chi^2$,
we can calculate the total $\chi^2$, which is plotted in
Fig.~\ref{sfig:chi2OmegaGWv2} as a function of $\Omega_{\rm GW}h_0^2 \sim
5.6 \times 10^{-6} (N_{\nu, \rm eff} - 3)$.

From this figure, we obtain the upper bound on the energy density of
the primordial GWs to be $\Omega_{\rm GW}h_0^2  < 1.8 \times 10^{-6}$
at 95$\%$ C.L. It is notable that this constraint is sensitive to both
the adiabatic and nonadiabatic components of radiation.
This constraint is shown in Fig.~\ref{fig:Illustration} as the gray dotted line.

\small

\bibliographystyle{utphys}
\bibliography{ref.bib}
\end{document}